\documentclass[%
 aip,
 amsmath,amssymb,
reprint, onecolumn 
]{revtex4-1}

\usepackage{graphicx}
\usepackage{dcolumn}
\usepackage{bm}

\usepackage[utf8]{inputenc}
\usepackage[T1]{fontenc}
\usepackage{mathptmx}
\usepackage{etoolbox}

\usepackage{cancel}
\usepackage{enumitem}

\makeatletter
\def\@email#1#2{%
 \endgroup
 \patchcmd{\titleblock@produce}
  {\frontmatter@RRAPformat}
  {\frontmatter@RRAPformat{\produce@RRAP{*#1\href{mailto:#2}{#2}}}\frontmatter@RRAPformat}
  {}{}
}%
\makeatother

\DeclareMathOperator{\erf}{erf}
\DeclareMathOperator{\erfc}{erfc}
\begin{document}
\def\opsim{\mathop{\sim}}
\def\opsimeq{\mathop{\simeq}}

\preprint{AIP/123-QED}

\title{Exact solutions for the probability density of various conditioned processes with an entrance boundary}
\author{Alain Mazzolo}
 \email{alain.mazzolo@cea.fr}
\affiliation{ 
$^{1}$Universit\'e Paris-Saclay, CEA, Service d'\'Etudes des R\'eacteurs et de Math\'ematiques Appliqu\'ees, 91191, Gif-sur-Yvette, France
}%


\date{\today}

\begin{abstract}
The probability density is a fundamental quantity for characterizing diffusion processes.  However, it is seldom known except in a few renowned cases, including Brownian motion and the Ornstein-Uhlenbeck process and their bridges, geometric Brownian motion, Brownian excursion, or Bessel processes. In this paper, we utilize Girsanov's theorem, along with a variation of the method of images, to derive the exact expression of the probability density for diffusions that have one entrance boundary. Our analysis encompasses numerous families of conditioned diffusions, including the Taboo process and Brownian motion conditioned on its growth behavior, as well as the drifted Brownian meander and generalized Brownian excursion. 
\end{abstract}

\maketitle

\newcommand{\E}{\mathrm{E}}
\newcommand{\Var}{\mathrm{Var}} 
\newcommand{\Cov}{\mathrm{Cov}}



\section{Introduction}
\label{sec_intro}
Effective stochastic processes resulting from the conditioning on specific events,
such as process survival or first passage to a target are of considerable theoretical
importance and have practical applications in various fields~\cite{ref_book_Karlin,ref_Baudoin,ref_Majumdar_Orland,ref_Adorisio,ref_Monthus_Mazzolo}.
However, the drift of conditioned processes can often feature singularities or have complex expressions. The exact analytical expression of the probability density of these diffusions, which provides valuable information about the process, remains mostly out of reach. Let us focus specifically on a one-dimensional diffusion process $\{X(t), 0 \le t \} $ that is characterized by its drift $\mu(x,t)$ and its variance $\sigma^2 = 1$. This diffusion $X(t)$ is driven by the Itô stochastic differential equation (SDE)
\begin{equation}
\label{eq_diffusion}
   \left\{
       \begin{aligned}
	  dX(t) & = \mu(X(t) ,t) dt + dW(t) \, , ~~ t \geq 0   \\       
	  X(0)  & = 0 \, ,
       \end{aligned}
   \right.
\end{equation}
where $W(t)$ is a standard Brownian motion (Wiener process). $\mathcal{L}. =    \mu(x,t)  \frac{\partial .}{\partial x}  + \frac{1}{2}  \frac{\partial^2 .}{\partial x^2}$ is the generator of the diffusion. To obtain the probability distribution $p(x,t)$ of this diffusion, it is necessary to solve the associated Fokker-Planck equation
\begin{equation}
\label{eq_fokker_planck_density_1D}
	 \frac{\partial p(x,t)}{\partial t} = -  \frac{\partial \left[\mu(x,t) p(x,t)\right]}{\partial x} + \frac{1}{2} \frac{\partial^2 p(x,t)}{\partial x^2} 
\end{equation}
with the appropriate boundary conditions. Several techniques are available to solve partial differential equations of this kind, such as
\begin{enumerate}[label=\arabic*)]
  \item Fourier transform~\cite{ref_book_Risken} 
  \item Laplace transform~\cite{ref_Frisch,ref_Mazzolo_Monthus_1,ref_Mazzolo_Monthus_2,ref_Mazzolo_Monthus_3} 
  \item The method of characteristic~\cite{ref_Kwok}
  \item The method of images, especially when absorption boundaries are present~\cite{ref_book_Redner,ref_intro_Redner}
  \item Eigenvalue expansion~\cite{ref_book_Karlin,ref_book_Risken,ref_book_Pavliotis}
  \item Path integral~\cite{ref_book_Risken} and Feynman-Kac technique~\cite{ref_Kac} 
  \item The Girsanov transform method~\cite{ref_book_Sarkka,ref_alain_arxiv}
\end{enumerate}
However, even with a range of techniques that span from classical analysis to purely probabilistic methods, the closed-form expression of the probability density is only known for a limited number of simple drifts. When there is an entrance boundary (or more), i.e. a boundary that cannot be reached from the interior of the state space~\cite{ref_book_Karlin}, the probability density's exact expression is generally unknown (with the notable exception of the Bessel process, whose probability density can be found in Borodin and Salminen's book~\cite{ref_book_Borodin}). This article examines six different drifts that share the characteristic of having an entrance boundary.  Specifically, we will determine the exact probability density related to the following drifts.
\begin{enumerate}[label=\Roman*)]
  \item $\mu_{I}(x) = -1/(a -x) ~~ \mathrm{with}~~ a>0$ (Taboo process)\\
   Entrance boundary at $a$ and state space $]-\infty,a[$.
  \item $\mu_{II}(x) = -\mu \coth\left[\mu(a-x)\right] ~~ \mathrm{with}~~ a>0$\\
   Entrance boundary at $a$ and state space $]-\infty,a[$.
  \item $\mu_{\alpha \beta}(x,t) = \alpha - \alpha \coth\left[\alpha(\alpha t + \beta - x \right)] ~~ \mathrm{with}~~ \alpha \in \mathbb{R}^* ~~ \mathrm{and}~~ \beta>0$\\
   Entrance boundary at $\alpha t + \beta $ and state space $]-\infty,\alpha t + \beta [$, for this process, both the entrance boundary and the state space are time-dependent.
     \item $\mu^*_{\alpha \beta}(x,t) = \alpha - \frac{1}{\alpha t + \beta - x} ~~ \mathrm{with}~~ \alpha < 0 ~~ \mathrm{and}~~ \beta>0$\\
   Entrance boundary and state space are the same as in case III.
    \item $\mu_{e}(x,t,X) = \frac{1}{T-t}\left( X \coth \left(\frac{X x}{T-t}\right) -x \right)$ with $T > t$.\\
    Brownian bridge between 0 and T ending at $X$ conditioned to stay positive i.e. the generalized Brownian excursion. Note that when $X=0$
    then $\mu_{e}(x,t,0) = \frac{1}{x} -\frac{x}{T-t}$ is the drift of the Brownian excursion between 0 and T. Entrance boundary at the origin and state space $]0,\infty[$.
    \item $\mu_{m}(x,t) = \mu + \frac{   2 \sqrt{\frac{2}{\pi (T-t)}} e^{-\frac{(x+\mu  (T-t))^2}{2 (t-T)}}  +2 \mu  e^{-2 \mu  x} \erfc\left(\frac{x - \mu  (T-t)}{\sqrt{2 (T-t)}}\right)}{1 + \erf\left(\frac{x + \mu  (T-t)}{\sqrt{2 (T-t)}}\right)-e^{-2 \mu  x} \erfc\left(\frac{x -\mu (T-t)}{\sqrt{2 (T-t)}}\right)}$ with $T > t$.\\
    Drifted Brownian meander between 0 and T. Entrance boundary and state space are the same as in case V.
\end{enumerate}
The first two drifts were recently obtained by conditioning a Brownian motion with constant drift $\mu$, with an absorbing boundary condition
at position $a$, toward full survival at the infinite horizon time~\cite{ref_Monthus_Mazzolo}. When the original drift is positive or equal to zero one obtains the drift $\mu_{I}(x)$, and when $\mu < 0$  the second drift $\mu_{II}(x)$. However, these two drifts have their roots in the mathematical literature. The first one corresponds to the celebrated Taboo process which is a Brownian motion that is conditioned to remain in the interval $]-\infty,a[$ forever. This process was introduced by Knight in 1969~\cite{ref_Knight}. The second can by found in Williams's seminal 1974 article~\cite{ref_Williams} where the author considers the diffusion with generator
\begin{equation}
\label{generator_Williams}
	 \mu \coth(\mu x)  \frac{d .}{d x}  + \frac{1}{2}  \frac{d^2 .}{dx^2} \, ,
\end{equation}
which, up to the constant $a$, corresponds to $\mu_{II}(x)$. The third and fourth drifts, $\mu_{\alpha \beta}(x,t)$ and $\mu^*_{\alpha \beta}(x,t)$ respectively, are both time- and space-dependent. They correspond to the conditioned diffusion which is achieved by ensuring that a Brownian motion remains below $\beta + \alpha t$ for all $t>0$. Specifically, the drift $\mu_{\alpha \beta}(x,t)$ corresponds to $\alpha \geq 0$, while $\mu^*_{\alpha \beta}(x,t)$ corresponds to $\alpha < 0$. These drifts as well as the last two (Brownian bridge conditioned to remain positive and drifted Brownian meander) are derived in the following section.\\

\noindent The goal of this article is to study in detail the six processes: $X_{I}(t)$ with drift $\mu_{I}(x)$, $X_{II}(t)$ with drift $\mu_{II}(x)$, $X_{\alpha \beta}(t)$ with drift $\mu_{\alpha \beta}(x,t)$, $X^*_{\alpha \beta}(t)$ with drift $\mu^*_{\alpha \beta}(x,t)$, $X_{e}(t)$ with drift $\mu_{e}(x,t,X)$ and $X_{m}(t)$ with drift $\mu_{m}(x,t)$. This will be accomplished by determining the exact expression of their probability density. To achieve this objective, we have developed a novel approach that merges Girsanov's theory with a variation of the image method.\\
The paper is structured as follows: In Section~\ref{sec2}, to obtain the exact expressions of the probability densities, we expose our method (Girsanov's theorem followed by a modified image method) by fully treating the case of the second drift $\mu_{II}(x)$. In the same section, this technique is also used to determine the exact expression of the probability densities for the five other drifts $\mu_{I}(x)$, $\mu_{\alpha \beta}(x,t)$, $\mu^*_{\alpha \beta}(x,t)$, $\mu_{e}(x,t,X)$ and $\mu_{m}(x,t)$.  Finally, Section~\ref{sec_Conclusion} presents some concluding remarks. In addition, Monte Carlo simulations illustrate some of our findings.

\section{Probability densities of the conditioned processes}
\label{sec2}
To properly characterize the six processes, namely $X_{I}(t)$, $X_{II}(t)$, $X_{\alpha \beta}(t)$, $X^*_{\alpha \beta}(t)$, $X_{e}(t)$ and $X_{m}(t)$, we will derive their probability densities in a closed-form.
However, solving the equation\eqref{eq_diffusion} with the drifts $\mu_{I}(x)$, $\mu_{II}(x)$, $\mu_{\alpha \beta}(x,t)$, $\mu^*_{\alpha \beta}(x,t)$, $\mu_{e}(x,t,X)$ or $\mu_{m}(x,t)$(or the corresponding Fokker-Planck equations) by a direct approach seems out of reach due to the nonlinearity of the drifts. Therefore, we adopt a different strategy by resorting to Girsanov's theorem whose derivation can be found, for example, in Karatzas and Shreve~\cite{ref_book_Karatzas} or \O ksendal~\cite{ref_book_Oksendal}. 
This theorem is often used to change the probability measures of stochastic differential equations and, in particular, we will utilize it to remove the initial drift hoping that in the driftless world, calculations become feasible. To provide a concrete example, let us consider the Itô SDE
\begin{equation}
\label{def_BM_with_drift_SDE_2}
   \left\{
       \begin{aligned}
	  dX(t) & = \mu(X(t),t) dt +  dW(t)  \, , ~~ t \geq 0 \\       
	  X(0)  & = x_0 \, ,
       \end{aligned}
   \right.
\end{equation}
\noindent where $W(t)$ is a standard Brownian motion. Girsanov's theorem states that the expectation of any function $h(X(t))$, where $X(t)$ is a solution of Eq.\eqref{def_BM_with_drift_SDE_2}, can be express as~\cite{ref_book_Sarkka}
\begin{equation}
\label{def_new_expectation}
	E[h(X(t))] = E[Z(t)h(\underbrace{x_0 + W(t)}_{\text{driftless process}})]  \, ,
\end{equation}
with
\begin{equation}
\label{def_Z}
   Z(t) =  e^{\textstyle\int_0^t \mu(x_0 + W(u),u) dW(u)  -\frac{1}{2} \int_0^t \mu(x_0 + W(u),u)^2 du}  \, .
\end{equation}
If the expression for $Z(t)$ is simple, then we can derive the closed-form of the probability density~\cite{ref_book_Sarkka}. We start by obtaining the probability density corresponding to the second drift $\mu_{II}(x)$. The probability density associated with the first drift will be obtained as the limit of the second probability density when $\mu \to 0$. The probability densities associated with the third drift $\mu_{\alpha \beta}(x,t)$ and the fourth $\mu^*_{\alpha \beta}(x,t)$  will be treated individually, along with the last two.\\

\subsection{Conditioned process with generator $\mathcal{L}_{II}. =  -\mu \coth\left[\mu(a-x)\right]   \frac{\partial .}{\partial x}  + \frac{1}{2}  \frac{\partial^2 .}{\partial x^2}$}
\label{subsec2}
\begin{figure}[h]
\centering
\includegraphics[width=4in,height=3.in]{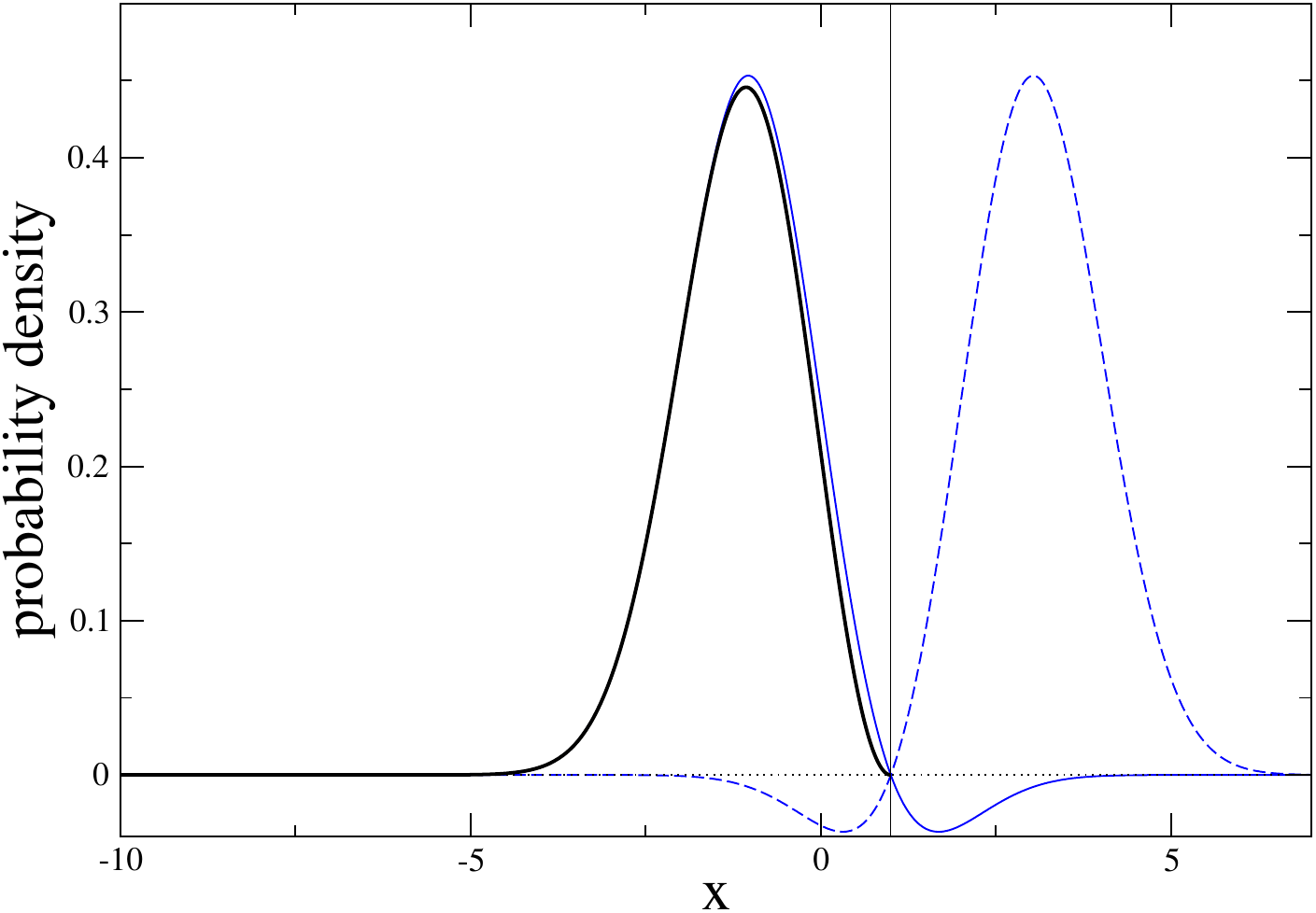}
\setlength{\abovecaptionskip}{15pt} 
\caption{$\tilde{p}_{II}(x,t)$ profile (blue curve) and the profile given by of its image $\tilde{p}_{II}(2 a - x,t)$ (blue dotted curve). The sum of both contributions (thick black curve) is the density probability function of the process $X_{II}(x,t)$ with drift $\mu_{II}(x) = -\mu \coth\left[\mu(a-x)\right]$ in the physical region $]-\infty,a[$. The parameters are $a=1$ (vertical line) and $\mu = -1$.}
\label{fig1}
\end{figure}

\noindent  As previously mentioned, we begin by examining the second diffusion process, $X_{II}(t)$, which is defined by its stochastic representation:
\begin{equation}
\label{def_BM_with_drift_SDE}
   \left\{
       \begin{aligned}
	  dX_{II}(t) & = -\mu \coth\left[\mu(a-X_{II}(t))\right] dt + dW(t) \, , ~~ t \geq 0   \\       
	  X_{II}(0)  & = 0 \, .
       \end{aligned}
   \right.
\end{equation}Reporting the expression of the second drift $\mu_{II}(x) = -\mu \coth\left[\mu(a-x)\right]$ into Eq.\eqref{def_Z} leads to (note that $x_0=0$ for the first four diffusions)
\begin{align}
\label{ZtdriftII}
	Z_{II}(t) & =  e^{\textstyle \int_0^t \mu_{II}(W(u)) dW(u)  -\frac{1}{2} \int_0^t \left(\mu_{II}(W(u))\right)^2 du}  \nonumber \\
	     & =  e^{\textstyle - \mu \int_0^t \coth\left[\mu(a-W(u))\right] dW(u)  -\frac{1}{2} \mu^2 \int_0^t \coth^2\left[\mu(a-W(u))\right] du} .
\end{align}
The stochastic integral can be evaluated thanks to Itô's formula
\begin{equation}
\label{itoformulamuII}
d\left[\log \left[ \sinh\left[\mu(a-W(u))\right] \right] \right] = - \mu \coth\left[\mu(a-W(u))\right] dW(u) -\frac{1}{2} \mu^2 \frac{du}{\sinh^2\left[\mu(a-W(u))\right]}  \, .
\end{equation}
Reporting this expression into Eq.\eqref{ZtdriftII} leads to
\begin{align}
\label{ZtdriftIIfinal}
	Z_{II}(t) & =  e^{{\big{[}} \textstyle  \log \left[ \sinh\left[\mu(a-W(u))\right] \right] {\big{]}}_0^t + \frac{1}{2} \mu^2 \textstyle \int_0^t \overbrace{\left\{ \frac{1}{\sinh^2\left[\mu(a-W(u))\right]} - \coth^2\left[\mu(a-W(u))\right] \right\} }^{= -1} du} 
	 \nonumber \\
			 & = \frac{\sinh\left[\mu(a-W(t))\right]}{\sinh\left[\mu a\right]} e^{- \frac{1}{2} \mu^2 t} \, .
\end{align}
Since this expression depends solely on the state of the Brownian motion at time $t$, the probability density can be evaluated explicitly~\cite{ref_book_Sarkka}. The probability density of the Brownian process $W(t)$ being $ \frac{1}{\sqrt{2 \pi t}} e^{- \frac{x^2}{2t}} $, in a perfect world, the probability density of the conditioned process $X_{II}(t)$ would be
\begin{equation}
\label{PDFwrongdrifII}
	\tilde{p}_{II}(x,t) = \frac{\sinh\left(\mu(a-x)\right)}{\sinh\left(\mu a\right)} e^{- \frac{1}{2} \mu^2 t}  \frac{1}{\sqrt{2 \pi t}} e^{- \frac{x^2}{2t}} \, . 
\end{equation}
From the previous equation it is straightforward to verify that $\tilde{p}_{II}(x,t)$ satisfies the Fokker-Planck equation
\begin{equation}
\label{Fokker-Planck-driftII}
	\partial_t \tilde{p}_{II}(x,t) -\frac{1}{2} \partial_{xx}^2  \tilde{p}_{II}(x,t) +\partial_{x}\left[ - \mu \coth\left[\mu(a-x)\right]  \tilde{p}_{II}(x,t) \right] = 0 \, .
\end{equation}
However, this is not the probability density of the process $X_{II}(t)$ for two reasons:
\begin{enumerate}[label=\alph*)]
\item The probability density is not normalized to unity
\begin{equation}
\label{normPDFwrongdrifII}
	\int_{-\infty}^a \tilde{p}_{II}(x,t) dx = \frac{ e^{ a \mu } \left(\erf\left(\frac{a+\mu  t}{\sqrt{2 t}}\right)+1\right)+e^{-a \mu } \left( \erfc\left(\frac{a-\mu  t}{\sqrt{2 t}}\right)-2 \right) }{4 \sinh\left(\mu a\right)}   > 1 \, .
\end{equation}
\item The outgoing current at the boundary $a$ is not equal to zero
\begin{equation}
\label{wrongcurrentdrifII}
	\tilde{j}(a,t) = - \frac{1}{2} \left[ \partial_{x} \tilde{p}_{II}(x,t) \right] \vert_{x=a} = \frac{\mu  e^{-\frac{a^2+\mu ^2 t^2}{2 t}}}{2 \sinh\left(\mu a\right) \sqrt{2 \pi t}} > 0 \, ,
\end{equation} 
meaning that the particle can cross the boundary, and this is not good since this result contradicts the assumption that $a$ is an entrance boundary, which as such, cannot be reached from inside the state space $]-\infty,a[$. 
\end{enumerate}
So, something peculiar happens with $\tilde{p}_{II}(x,t)$ when Girsanov's theorem is not used with enough care.
Indeed, the Brownian motion exists on the real line at any time $t>0$, whereas $X_{II}(x,t)$ is only defined on the interval $]-\infty,a[$. As a result, the two probability measures associated with the processes are not equivalent probability measures. In such a case, the change of probability measure imposed by Girsanov's theorem leads to a new "density function" which is in general not a probability density function (see for instance~\cite{ref_book_Mikosch} for more detail). That point being clarified, $\tilde{p}_{II}(x,t)$  is a solution of the Fokker-Planck equation~Eq.\eqref{Fokker-Planck-driftII} and besides it is normalized on the whole (unphysical) real line
\begin{equation}
\label{normPDFdrifII_on_R}
	\int_{-\infty}^{\infty} \tilde{p}_{II}(x,t) dx = 1 \, .
\end{equation}
Looking at the shape of $\tilde{p}_{II}(x,t)$ (see Fig.~\ref{fig1})
strongly suggests that the probability density function $p_{II}(x ,t)$ should be given as the sum of $\tilde{p}_{II}(x ,t)$ and its symmetrical image with respect to the boundary $a$, i.e. $p_{II}(x ,t) = \tilde{p}_{II}(x ,t) + \tilde{p}_{II}(2a - x ,t)$  (note that a positive image is added, not a negative one as in the standard image method with an absorbing boundary condition~\cite{ref_book_Redner, ref_intro_Redner}). We claim that this superposition gives a resultant probability density function that solves the original problem in the interval $]-\infty,a[$. Indeed, we get
\begin{align}
\label{PDFdrifII}
	 p_{II}(x,t) & = \tilde{p}_{II}(x ,t) + \tilde{p}_{II}(2a - x ,t) \nonumber \\
	 	      & = \frac{\sinh\left(\mu(a-x)\right)}{\sinh\left(\mu a\right)} e^{- \frac{1}{2} \mu^2 t}  \frac{1}{\sqrt{2 \pi t}} e^{- \frac{x^2}{2t}} + \frac{\sinh\left(\mu(-a+x)\right)}{\sinh\left(\mu a\right)} e^{- \frac{1}{2} \mu^2 t}  \frac{1}{\sqrt{2 \pi t}} e^{- \frac{(2 a -x)^2}{2t}} \nonumber \\
		      & = \frac{\sinh\left(\mu(a-x)\right)}{\sinh\left(\mu a\right)} e^{- \frac{1}{2} \mu^2 t}  \frac{1}{\sqrt{2 \pi t}} \left(e^{- \frac{x^2}{2t}} - e^{- \frac{(2 a -x)^2}{2t}}  \right) \, .
\end{align}
First, $p_{II}(x,t)$ is non-negative on $]-\infty,a[ \times [0,\infty[$. Furthermore, since $p_{II}(x,t)$ is the sum of two functions that fulfill~Eq.\eqref{Fokker-Planck-driftII}, it is also a solution of the Fokker-Planck equation. Moreover, direct calculation shows that $p_{II}(x,t)$ is properly normalized
\begin{equation}
\label{normPDFdrifII}
	\int_{-\infty}^{a} p_{II}(x,t) dx = 1 \, .
\end{equation}
In addition, the current is 
\begin{align}
\label{currentdrifII}
	j_{II}(x,t) & = - \frac{1}{2} \partial_{x} p_{II}(x,t) \nonumber \\
	       & = \frac{1}{2} \frac{e^{- \frac{1}{2} \mu^2 t}}  {\sinh\left(\mu a\right) \sqrt{2 \pi t^3}} \left[\mu t \cosh\left[\mu(a-x)\right]  \left(e^{- \frac{x^2}{2t}} - e^{- \frac{(2 a -x)^2}{2t}}  \right)  
	       \sinh\left(\mu(a-x)\right)\left( x e^{- \frac{x^2}{2t}} + (2 a -x) e^{- \frac{(2 a -x)^2}{2t}} \right) \right] \, ,
\end{align} 
and thus $j_{II}(a,t) = 0$ as it should. This allows us to conclude that $p_{II}(x,t)$, given by~Eq.\eqref{PDFdrifII}, is the probability density of the process $X_{II}(t)$.\\
The mean behavior and variance of the process can be obtained directly from this probability density. For the average we obtain
\begin{equation}
\label{XmeantypeII}
	\E[X_{II}(t)] = \frac{e^{ a \mu } \left(a-(a+\mu  t) \erf\left(\frac{a+\mu  t}{\sqrt{2 t}}\right)\right) +e^{-a \mu } \left(-a+(a-\mu  t) \erf\left(\frac{a-\mu  t}{\sqrt{2 t}}\right)\right)}{2 \sinh(a \mu )}  \, ,
\end{equation}
and for the variance
\begin{align}
\label{XvartypeII}
	\Var[X_{II}(t)]  = & -\frac{e^{-2 a \mu }}{4 \, {\sinh}^2(a \mu )}   \left(e^{2 a \mu } \left(a-(a+\mu  t) \erf\left(\frac{a+\mu  t}{\sqrt{2 t}}\right)\right)+(a-\mu  t) \erf\left(\frac{a-\mu  t}{\sqrt{2 t}}\right)-a\right)^2  \nonumber \\
	       & + \frac{\coth (a \mu )-1}{2}  \left(2 a e^{2 a \mu } (a+\mu  t) \erfc\left(\frac{a+\mu  t}{\sqrt{2 t}}\right)-2 a (a-\mu  t) \erfc\left(\frac{a-\mu  t}{\sqrt{2 t}}\right)
	        +t \left(e^{2 a \mu }-1\right) \left(\mu ^2 t+1\right)\right) \, .
\end{align}
For large times $t$, one can compute the leading order of the two previous expressions 
\begin{equation}
\label{Xmean_and_Var_asymptotic_typeII}
   \left\{
       \begin{aligned}
        \E[X_{II}(t)]      & \underset{t \to \infty}{\sim\,} \mu \, t  \\
        \Var[X_{II}(t)]    & \underset{t \to \infty}{\sim\,} t  \, .
       \end{aligned}
   \right.
\end{equation}
Thus, in the long-time limit, the process $X_{II}(t)$ behaves like a Brownian motion with negative drift $\mu$. Loosely speaking, for large times, the process does not feel the boundary $a$ since it wanders to minus infinity.

\subsection{Conditioned process with generator $\mathcal{L}_{I}. = - \frac{1}{(a-x)} \frac{\partial .}{\partial x}  + \frac{1}{2}  \frac{\partial^2 .}{\partial x^2}$ (Taboo process)} 
\label{subsec3}
The drift $\mu_{I}(x) = -1/(a -x)$, which corresponds to the drift of the taboo process~\cite{ref_Knight,ref_Pinsky,ref_Mazzolo_Taboo}, can be obtained as the limit
\begin{equation}
	\lim_{\mu \to 0} -\mu \coth\left[\mu(a-x)\right] = -\frac{1}{a -x} \, ,
\end{equation}
meaning that the taboo process can be obtained from the second process $X_{II}(t)$ in the limit as $\mu$ tends to zero. Thus, the probability density of the taboo process can be derived directly from the probability density of the second process and is given by
\begin{equation}
	p_{I}(x,t) = \lim_{\mu \to 0}  p_{II}(x,t) = \frac{(a-x) }{a \sqrt{2 \pi t}} \left(e^{-\frac{x^2}{2 t}} -  e^{-\frac{(2 a-x)^2}{2 t}} \right) \, .
\end{equation}
This result can be found in~\cite{ref_Knight} where it is obtained in this specific case by studying the joint densities of $B(t)$ given $\max_{0 \leq t \leq T} B(t) < a$. 
It can be easily verified that $p_{I}(x,t)$ satisfies the Fokker-Planck equation
\begin{equation}
\label{Fokker-Planck-driftI}
	\partial_t p_{I}(x,t) -\frac{1}{2} \partial_{xx}^2  p_{I}(x,t) +\partial_{x}\left[ - \frac{1}{(a-x)} p_{I}(x,t) \right] = 0 \, .
\end{equation}
Also the function is positive on  $]-\infty,a[ \times [0,\infty[$, normalized to unity, and the current
\begin{equation} 
\label{currentdrifI}
 j_{I}(x,t)  = - \frac{1}{2} \partial_{x} p_{I}(x,t) = \frac{x (a-x) e^\frac{-x^2}{2 t}+(a-x) (2 a-x) e^{-\frac{(2 a-x)^2}{2 t}}+ t e^{-\frac{x^2}{2 t}}- t e^{-\frac{(2 a-x)^2}{2 t}}}  {2 a \sqrt{2 \pi t^3}}  \, ,
\end{equation}
is equal to zero on the boundary $a$. Thanks to the exact probability density expression, we can easily calculate the expectation and the variance of the taboo process. We get respectively
\begin{equation}
\label{XmeantypeI}
	\E[X_{I}(t)] = a -\sqrt{\frac{2 t}{\pi }}  e^{- \frac{a^2}{2 t}} -\frac{\left(a^2+t\right) \erf\left(\frac{a}{\sqrt{2 t}}\right)}{a}  \, ,
\end{equation}
and
\begin{align}
\label{XvartypeI}
	\Var[X_{I}(t)] & = a^2 + \left(3-\frac{2}{\pi } e^{-\frac{a^2}{t}}\right) t - \left(1+ \frac{t}{a^2} \right) \erf\left(\frac{a}{\sqrt{2 t}}\right) \left(2 a \sqrt{\frac{2 t}{\pi }}   e^{-\frac{a^2}{2 t}} + \left(a^2+t\right) \erf\left(\frac{a}{\sqrt{2 t}}\right)  \right) \, ,
\end{align}
as well as their asymptotic behaviors for large times
\begin{equation}
\label{Xmean_and_Var_asymptotic_typeI}
   \left\{
       \begin{aligned}
        \E[X_{I}(t)]      & \underset{t \to \infty}{\sim\,} -2 \sqrt{\frac{2  t}{\pi }}  \\
        \Var[X_{I}(t)]    & \underset{t \to \infty}{\sim\,} \left(3-\frac{8}{\pi }\right) t \, .
       \end{aligned}
   \right.
\end{equation}
Thus, for large times, the average behavior of the taboo process is nearer to the boundary $a$ than the $X_{II}(t)$ process.

\subsection{Conditioned process with generator $\mathcal{L_{\alpha \beta}}. = \alpha \left(1 - \coth\left[\alpha(\alpha t + \beta -x)\right] \right)  \frac{\partial .}{\partial x}  + \frac{1}{2}  \frac{\partial^2 .}{\partial x^2}$}
\label{subsec4}
First, we will explain where this generator comes from. A brief overview of the Doob conditioning technique will then be provided and applied to the Brownian motion $B(t)$ conditioned on its growth behavior. More specifically, we will consider the conditioned process $X_{\alpha \beta}(t)$ defined by:
\begin{align}
	X_{\alpha \beta}(t) = B(t) & \mathrm{~~constrained~such~that~~} B(t) \leq \beta + \alpha t  \mathrm{~~for~all~~} t>0 \\ \nonumber
	                           & \mathrm{~~with~~} \alpha > 0 \mathrm{~~and~~} \beta > 0 \,  ,
\end{align}
and show that this process has a generator given by $\mathcal{L_{\alpha \beta}}$. 
Imposing constraints on the Brownian motion is commonly achieved through Doob's {\it{h}}-transform~\cite{ref_Doob}, as detailed in Chapter 15 of Karlin and Taylor's book~\cite{ref_book_Karlin} and outlined from a physicist's perspective in the recent article~\cite{ref_Majumdar_Orland}. The key ingredient of Doob's method requires the calculation of the conditioned probability $\pi(x,t)$ that from the state value $x$ at time $t$, the sample path of the process satisfies the desired constraint at time $T$ ($T$ being finite or infinite). When this quantity is known, Doob's technique have been successfully applied to various kind of conditioned processes~\cite{ref_book_Karlin,ref_Baudoin,ref_Majumdar_Orland,ref_Szavits,ref_Chetrite,ref_Larmier}. Let us consider a diffusion process $\{X(t), 0 \le t \le T\}$ characterized by a drift $\mu(x)$ and a variance $\sigma^2(x)$. Consequently, the process $X(t)$ satisfies the stochastic differential equation
\begin{equation}
\label{diffusion_general}
  dX(t) = \mu(X(t)) dt+ \sigma(X(t)) dW(t) \, ,
\end{equation}
\noindent with the initial value $X(0) = x_0$. Now, let $\{X^*(t), 0 \le t \le T\}$ be the process conditioned on an event between the two times 0 and $T$: for instance, for a Brownian bridge the constraint is the event $\{X(T)=0\}$,  while for the present study it is the event $\{X(t) \leq \beta + \alpha t  \mathrm{~~for~all~~} t>0\}$. Then, according to Doob's conditioning, the constrained process is characterized by the drift $\mu^*(x,t)$ and variance $\sigma^{*2}(x,t)$ respectively given by~\cite{ref_book_Karlin}:
\begin{equation}
\label{Doob_general}
   \left\{
       \begin{aligned}
       \sigma^*(x,t) & = \sigma(x) \, ,  \\
       \mu^*(x,t)    & = \mu(x) + \frac{ \sigma^2(x)}{\pi(x,t)}   \frac{\partial \pi(x,t)}{\partial x}  \, .
       \end{aligned}
   \right.
\end{equation}
Thus, all we need to fully characterize the Brownian motion conditioned on its growth behavior is the conditioned probability defined by:
\begin{equation}
\label{conditioned_probability_definition}
	\pi(x,t) = \mathrm{Pr}[B(u) \leq \alpha u + \beta \mathrm{~~for~all~~} u \geq t \vert B(t)=x] 	
\end{equation}
assuming that $ x \leq \alpha t + \beta$. From~\cite{ref_book_Karlin_I} we know that for  $\alpha > 0$
\begin{equation}
\label{conditioned_probability_KT}
	\mathrm{Pr}[B(t) \leq \alpha t + \beta \mathrm{~~for~all~~} t \vert B(0)=x] = 1- e^{-2 \alpha (\beta -x)} \, ,
\end{equation}
therefore by a simple shift in time, we get the desired conditioned probability
\begin{equation}
\label{conditioned_probability}
	\pi(x,t) =  1- e^{-2 \alpha (\alpha t + \beta -x)} 	.
\end{equation}
Reporting this expression into~Eq.\eqref{Doob_general} leads to
\begin{equation}
\label{mu_alpha_beta}
       \mu_{\alpha \beta}(x,t)    = \alpha - \alpha  \coth\left[\alpha(\alpha t + \beta -x)\right] 
\end{equation}
which is the drift corresponding to the generator $\mathcal{L_{\alpha \beta}}$. Our expression corrects the expression of $\mu^*(x)$ given in Chapter 15 (p. 272) of Reference~\cite{ref_book_Karlin}, which contains a typo and is independent of time. It should also be noted that this result, which is obtained from Doob conditioning, is only valid for $\alpha > 0$, otherwise for $\alpha <0$ the probability $\pi(x,t)$ is strictly equal to zero (since the Brownian motion almost surely reaches the downward sloping line $\beta - \vert \alpha \vert t$). Hence, the formula~\eqref{Doob_general} cannot be directly applied. This case will be discussed in detail in the upcoming subsection~\ref{subsec5}. However, from a formal point of view, there are no limitations to study the process whose generator is $\mathcal{L_{\alpha \beta}}$ with $\alpha \in \mathbb{R^*}$. This will be our focus in the following section.\\

\begin{figure}
\centering
\includegraphics[width=4in,height=3.in]{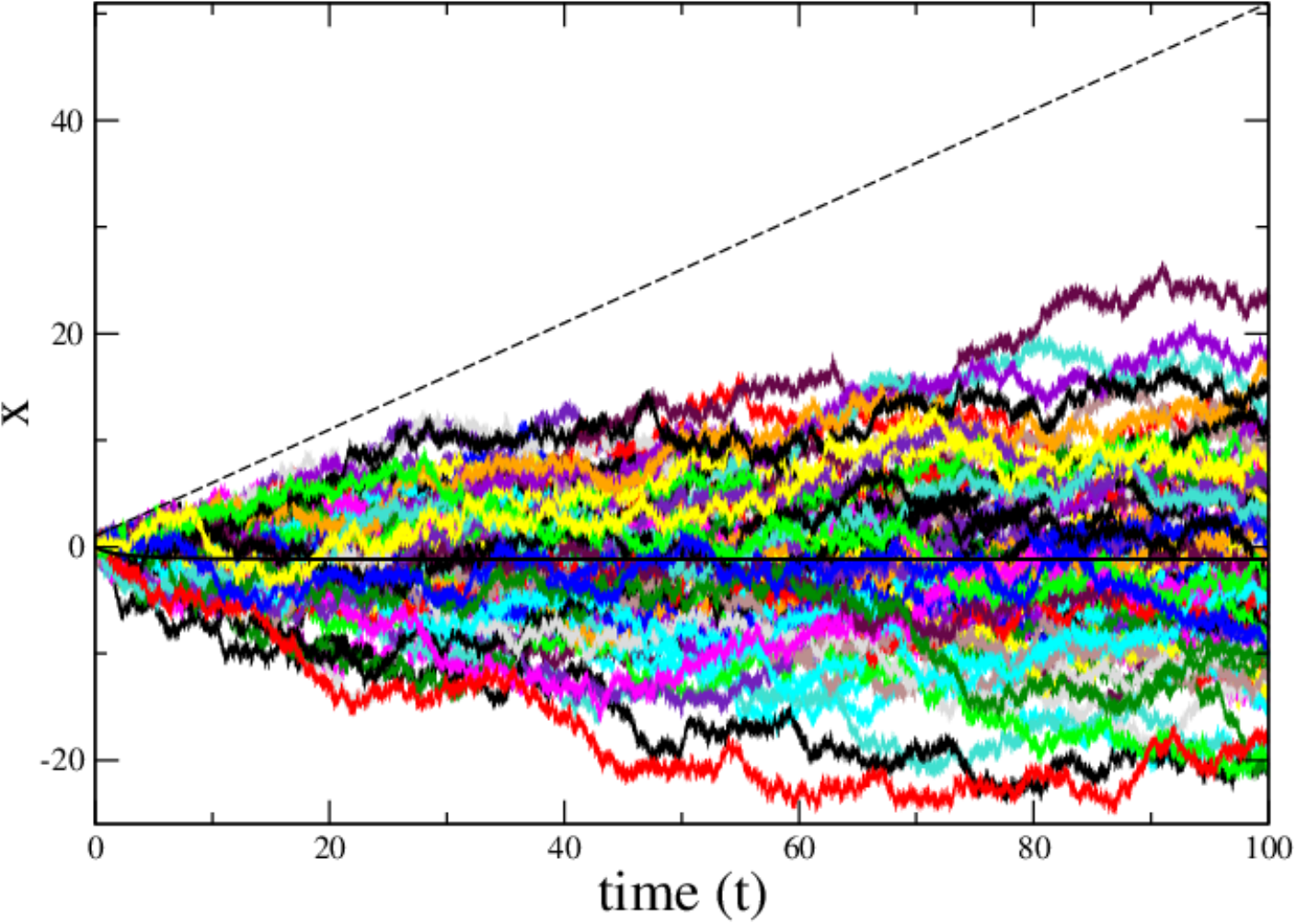}
\setlength{\abovecaptionskip}{15pt} 
\caption{A sample of 100 diffusions for the conditioned process $X_{\alpha \beta}(t)$ with parameters $\alpha = 1/2$ and $\beta = 1$. The time step used in the
discretization is $dt = 10^{-3}$. All trajectories generated with different noise histories are
statistically independent. The thick black curve is the average profile of the stochastic
process given by~Eq.\eqref{Xmeantype_alpha_beta}. Observe that at large times the process no longer feels the boundary and evolves almost freely.}
\label{fig2}
\end{figure}

\begin{figure}
\centering
\includegraphics[width=4in,height=3.in]{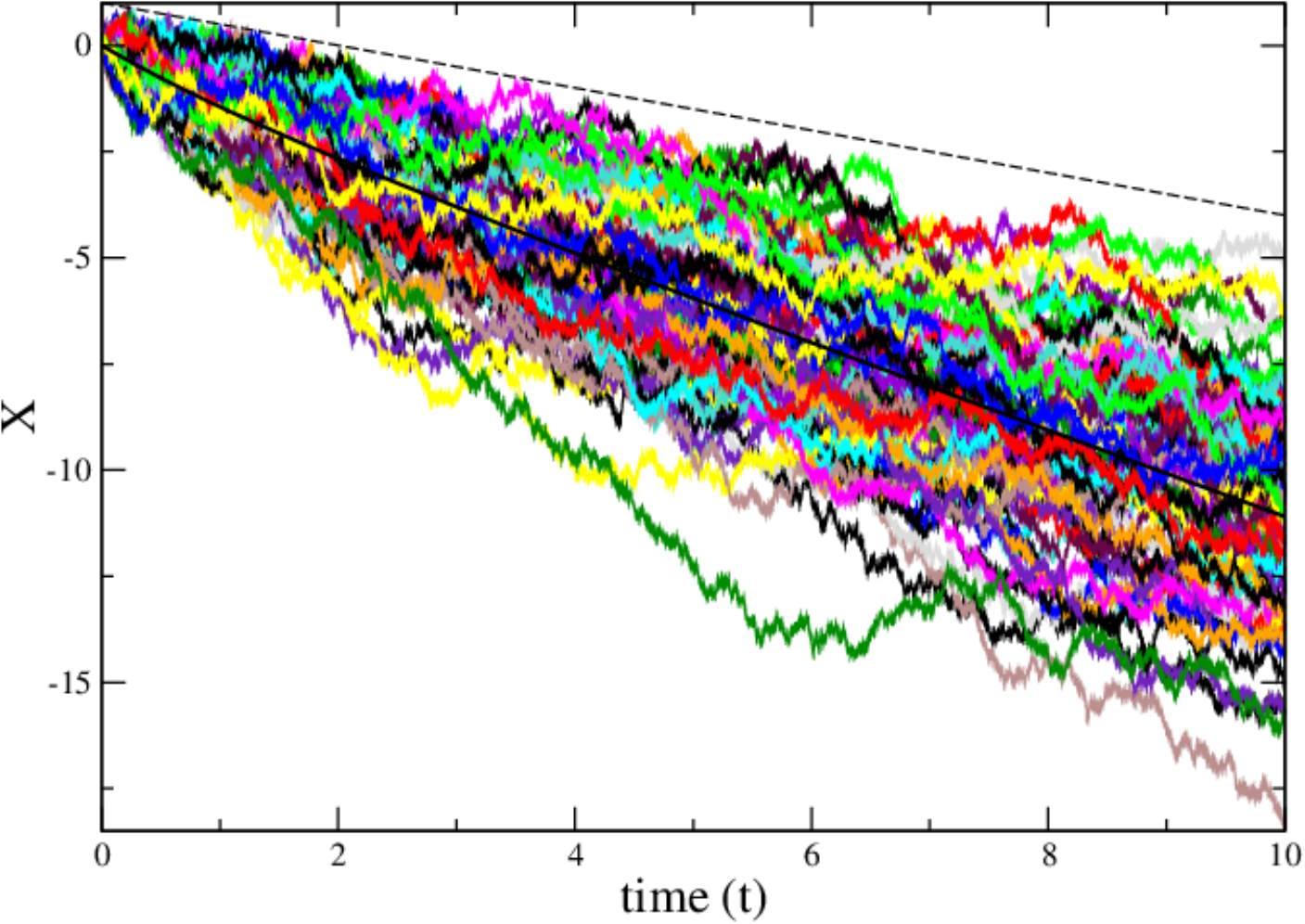}
\setlength{\abovecaptionskip}{15pt} 
\caption{A sample of 100 diffusions for the conditioned process $X_{\alpha \beta}(t)$ with parameters $\alpha = -1/2$ and $\beta = 1$. The time step used in the
discretization is $dt = 10^{-4}$. All trajectories generated with different noise histories are
statistically independent. The thick black curve is the average profile of the stochastic process given by~Eq.\eqref{Xmeantype_alpha_beta}.}
\label{fig3}
\end{figure}

\noindent As in the previous case we proceed by inserting the expression of $\mu_{\alpha \beta}(x,t)$ into Eq.\eqref{def_Z}, we get
\begin{align}
\label{Ztdrift_alpha_beta}
	Z_{\alpha \beta}(t) & =  e^{\textstyle \int_0^t \mu_{\alpha \beta}(W(u),u) dW(u)  -\frac{1}{2} \int_0^t \left( \mu_{\alpha \beta}(W(u),u) \right)^2 du}   \\
	     & =  e^{\textstyle \alpha \int_0^t  \left(1 - \coth\left[\alpha(\alpha t + \beta -W(u))\right] \right) dW(u)  -\frac{\alpha^2}{2} \int_0^t  \left(1 - \coth\left[\alpha(\alpha t + \beta -W(u))\right] \right)^2 du} \nonumber \, .
\end{align}
The first integral can be evaluated by applying Itô's formula
\begin{align}
\label{itoformula_alpha_beta}
	d\left\{\alpha W(u) + \log \left[ \sinh\left[\alpha(\alpha u + \beta -W(u))\right] \right] \right\} &=	\alpha^2  \coth\left[\alpha(\alpha u + \beta -W(u))\right] du  \nonumber \\
	 & + \alpha \left(1 - \coth\left[\alpha(\alpha u + \beta -W(u))\right] \right) dW(u)  \nonumber \\
	 & - \frac{\alpha^2}{2} \frac{du}{\sinh^2\left[\alpha(\alpha u + \beta -W(u))\right]} \, ,
\end{align}
which reads
\begin{align}
\label{First_intergral_driftdoobmuneglambda}
	\alpha \int_0^t \left(1 - \coth\left[\alpha(\alpha u + \beta -W(u))\right] \right) dW(u) & = \big[\alpha W(u) + \log \left[ \sinh\left[\alpha(\alpha u + \beta -W(u))\right] \right] \big]_0^t \nonumber \\
				& - \alpha^2 \int_0^t \coth\left[\alpha(\alpha u + \beta -W(u))\right] du \nonumber \\
				& + \frac{\alpha^2}{2}  \int_0^t \frac{dt}{\sinh^2\left[\alpha(\alpha u + \beta -W(u))\right]} \, .
\end{align}
Reporting this expression in Eq.\eqref{Ztdrift_alpha_beta} yields
\begin{align}
\label{Ztdrift_alpha_beta_final}
	Z_{\alpha \beta}(t) & = e^{\textstyle \alpha W(t) + \log \left[ \sinh\left[\alpha(\alpha t + \beta -W(t))\right] \right] - \log \left[ \sinh\left[\alpha \beta \right] \right]  }  \nonumber \\
	     & \times   e^{ - \alpha^2 \int_0^t \overbrace{ \left( \scriptstyle  \coth\left[\alpha(\alpha t + \beta -W(t))\right] - \frac{1}{2 \sinh^2\left[\alpha(\alpha t + \beta -W(t))\right]}+ \frac{1}{2}  \left(1 - \coth\left[\alpha(\alpha t + \beta -W(t))\right] \right)^2 \right)}^{=1 } dt }
	    \nonumber  \\
	     & =  \frac{\sinh\left[\alpha(\alpha t + \beta -W(t))\right]}{\sinh\left[\alpha \beta \right]} 	 e^{\textstyle \alpha W(t) -\alpha^2 t}     .
\end{align}
As in the first case studied, this expression only relies on the state of the Brownian motion at time $t$. Therefore a probability density $\tilde{p}_{\alpha \beta}(x,t)$ of the conditioned process -which as in case II is not a probability function since the probability measures associated with the Brownian motion and the Brownian motion conditioned on its growth behavior are not equivalent probability measures- can be evaluated explicitly, we obtain
\begin{equation}
\label{PDFwrongdrif_alpha_beta}
	\tilde{p}_{\alpha \beta}(x,t) =  \frac{\sinh\left(\alpha(\alpha t + \beta -x)\right)}{\sinh\left(\alpha \beta \right)} e^{ \alpha x -\alpha^2 t} \frac{1}{\sqrt{2 \pi t}} e^{- \frac{x^2}{2t}} \, ,
\end{equation}
which by construction satisfies the Fokker-Planck equation
\begin{equation}
\label{Fokker-Planck-drift_alpha_beta}
	\partial_t \tilde{p}_{\alpha \beta}(x,t) -\frac{1}{2} \partial_{xx}^2  \tilde{p}_{\alpha \beta}(x,t) +\partial_{x}\left[ \alpha \left(1 - \coth\left[\alpha(\alpha t + \beta -x)\right] \right)  \tilde{p}_{\alpha \beta}(x,t) \right] = 0 \, .
\end{equation}
To obtain the true probability density, we apply the variant of the image method presented in Sec.\ref{subsec2} by adding the density function of a symmetrical particle with respect to the boundary at time $t$ and thus located at $2 (\beta +\alpha t)-x$. This position is time-dependent, since the boundary $\beta +\alpha t$ varies with time. We get
\begin{align}
\label{PDFdrif_alpha_beta}
	 p_{\alpha \beta}(x,t) & = \tilde{p}_{\alpha \beta}(x ,t) + \tilde{p}_{\alpha \beta}(2 (\beta +\alpha t)-x ,t) \nonumber \\
	 	      & = \frac{\sinh\left(\alpha(\alpha t + \beta -x)\right)}{\sinh\left(\alpha \beta \right)} e^{\alpha x -\alpha^2 t} \frac{1}{\sqrt{2 \pi t}} e^{- \frac{x^2}{2t}}  \nonumber \\
	 	      & + \frac{\sinh\left(\alpha(\alpha t + \beta -(2 (\beta +\alpha t)-x))\right)}{\sinh\left(\alpha \beta \right)} e^{\alpha (2 (\beta +\alpha t)-x)) -\alpha^2 t} \frac{1}{\sqrt{2 \pi t}} e^{- \frac{(2 (\beta +\alpha t)-x))^2}{2t}} \nonumber \\
		      & =  \frac{\sinh\left(\alpha(\alpha t + \beta -x)\right)}{\sinh\left(\alpha \beta \right)} e^{-\alpha ^2 t+\alpha  (x-2 \beta )-\frac{4 \beta ^2+x^2}{2 t}} \frac{1}{\sqrt{2 \pi t}} \left( e^{\frac{2 \beta  x}{t}}-e^{\frac{2 \beta  (\beta +\alpha  t)}{t}} \right) \, ,
\end{align}
with
\begin{equation}
\label{normPDFdrif_alpha_beta}
	\int_{-\infty}^{\beta +\alpha t} p_{\alpha \beta}(x,t) dx = 1 \, ,
\end{equation}
and 
\begin{equation} 
\label{currentdrif_alpha_beta}
     j_{\alpha \beta}(\beta +\alpha t,t)  = - \frac{1}{2} \partial_{x} p_{\alpha \beta}(x,t) \Big{|}_{x = \beta +\alpha t}= 0  \, .
\end{equation}
Besides $p_{\alpha \beta}(x,t) \geq 0 $ on the domain $D=\{(x,t)\vert  ~ x<\beta + \alpha t , ~ t \geq 0\}$,  therefore $p_{\alpha \beta}(x,t) $, given by~Eq.\eqref{PDFdrif_alpha_beta}, is the density probability of the conditioned process whose generator is 
$\mathcal{L_{\alpha \beta}}$. \\
\noindent With the closed-form expression of the probability density at our disposal, we can obtain the expectation and the variance of the conditioned process. We get, respectively
\begin{equation}
\label{Xmeantype_alpha_beta}
	\E[X_{\alpha \beta}(t)] = \frac{e^{-\alpha  \beta }} {2 \sinh(\alpha  \beta )} \left(-2 \beta +(\beta -\alpha  t) \text{erfc}\left(\frac{\alpha  t-\beta }{\sqrt{2 t}}\right)+e^{2 \alpha  \beta } (\beta + \alpha  t  )  \text{erfc}\left(\frac{\alpha  t +\beta}{\sqrt{2 t}}\right)\right)  \, ,
\end{equation}
and
\begin{align}
\label{Xvartype_alpha_beta}
	& \Var[X_{\alpha \beta}(t)] = -   \frac{e^{-2 \alpha  \beta }}{4  \sinh^2(\alpha  \beta )} \left(-2 \beta +(\beta -\alpha  t) \erfc\left(\frac{\alpha  t-\beta }{\sqrt{2 t}}\right)+e^{2 \alpha  \beta } (\beta +\alpha  t) \text{erfc}\left(\frac{\beta +\alpha  t}{\sqrt{2 t}}\right)\right)^2 \nonumber \\
	       & +  \frac{e^{-\alpha  \beta }}{2 \sinh(\alpha  \beta )} \left(-4 \beta ^2+2 \left(\beta ^2-\alpha ^2 t^2\right) \erfc\left(\frac{\alpha  t-\beta }{\sqrt{2 t}}\right)+e^{2 \alpha  \beta } \left(2 (\beta +\alpha  t)^2 \erfc\left(\frac{\beta +\alpha  t}{\sqrt{2 t}}\right)+t\right)-t\right) \, ,
\end{align}
and their asymptotic behaviors for large times
\begin{equation}
\label{Xmean_and_Var_asymptotic_type_alpha_beta}
   \left\{
       \begin{aligned}
        \E[X_{\alpha \beta}(t)]      & \underset{t \to \infty}{\sim\,}   \left\{
                                      \begin{aligned} 
                                               & \beta -\beta  \coth (\alpha  \beta ) & \mathrm{~if~~} \alpha > 0 \\
                                               & 2 \alpha  t  & \mathrm{~if~~} \alpha< 0  \\
                                      \end{aligned}     
                                                              \right.    
       \\
        \Var[X_{\alpha \beta}(t)]    & \underset{t \to \infty}{\sim\,} t \, .
       \end{aligned}
   \right.
\end{equation}
Depending on the sign of $\alpha$, the long-term behavior of the process is radically different. For positive $\alpha$, corresponding to an increasing domain over time, the mean value of the process approaches a constant and evolves almost freely, as shown in Fig.~\ref{fig2}. For negative $\alpha$, corresponding to a shrinking domain over time, the mean value of the process evolves towards $2 \alpha t$, i.e. roughly 2 times the value of the boundary at time t (see Fig.~\ref{fig3}).

\subsection{Conditioned process with generator $\mathcal{L^*_{\alpha \beta}}. =  \left(\alpha - \frac{1}{\alpha t + \beta -x} \right)  \frac{\partial .}{\partial x}  + \frac{1}{2}  \frac{\partial^2 .}{\partial x^2}$}
\label{subsec5}
In this paragraph, we again consider the Brownian motion $B(t)$ conditioned to remain below a straight line, but this time with a decreasing slope. More formally, we consider the conditioned process $X^*_{\alpha \beta}(t)$ defined by:
\begin{align}
	X^*_{\alpha \beta}(t) = B(t) & \mathrm{~~constrained~such~that~~} B(t) \leq \beta + \alpha t  \mathrm{~~for~all~~} t>0 \\ \nonumber
	                           & \mathrm{~~with~~} \alpha < 0 \mathrm{~~and~~} \beta > 0 \, .
\end{align}
If we now consider the Brownian motion $\tilde{B}(t) = B(t) - \alpha t$ with a positive uniform drift $-\alpha$, then the previous condition is equivalent to $\tilde{B}(t)$ remaining forever below the positive boundary $\beta$. This case is discussed in the recent article~\cite{ref_Monthus_Mazzolo} where the conditioning leads to a Taboo-like stochastic process driven by the SDE:
\begin{equation}
\label{SDE_aux}
       dX^*(t) = -\frac{1}{\beta - X^*(t)} dt + dW(t)
\end{equation}
and thus for the original conditioned process
\begin{equation}
\label{SDE_final}
       dX^*_{\alpha \beta}(t) = \left(\alpha -\frac{1}{\beta +\alpha t- X^*_{\alpha \beta}(t) } \right) dt + dW(t) \, ,      
\end{equation}
whose generator corresponds to $\mathcal{L^*_{\alpha \beta}}.$\\

\begin{figure}[h]
\centering
\includegraphics[width=3.5in,height=2.8in]{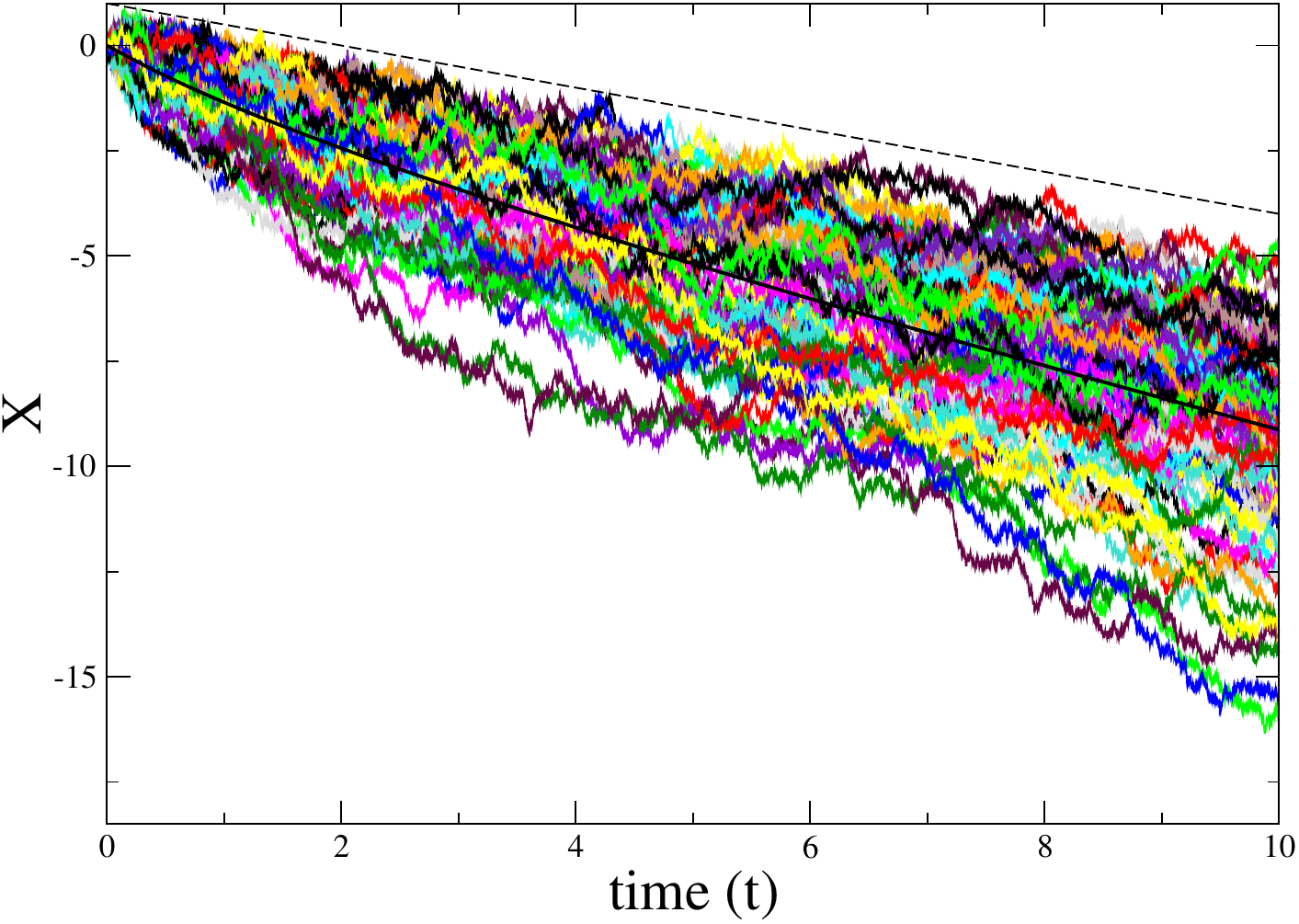}
\setlength{\abovecaptionskip}{15pt} 
\caption{A sample of 100 diffusions for the conditioned process $X^*_{\alpha \beta}(t)$ with parameters $\alpha = -1/2$ and $\beta = 1$. The time step used in the
discretization is $dt = 10^{-4}$. All trajectories generated with different noise histories are
statistically independent. The thick black curve is the average profile of the stochastic
process given by~Eq.\eqref{Xmeantype_alpha_beta*}. Observe that the process $X^*_{\alpha \beta}(t)$ resulting from Doob conditioning remains globally  closer to the boundary than process $X_{\alpha \beta}(t)$ studied in Sec.\ref{subsec4} and presented in Fig.~\ref{fig3}.}
\label{fig4}
\end{figure}

With the help of the transformation: $W(t) \mapsto \alpha W(t) + \log[\beta + \alpha t -W(t)]$, the probability density can be obtained using the technique presented in the previous paragraphs. In this vein, we obtain
\begin{equation}
\label{PDFdrif_alpha_beta*}
	 p^*_{\alpha \beta}(x,t) = -\frac{(\alpha t + \beta -x)}{ \beta } e^{-\frac{\alpha ^2}{2} t+\alpha  (x-2 \beta )-\frac{4 \beta ^2+x^2}{2 t}} \frac{1}{\sqrt{2 \pi t}} \left( e^{\frac{2 \beta  x}{t}}-e^{\frac{2 \beta  (\beta +\alpha  t)}{t}} \right) \, .
\end{equation}
With the closed-form expression of the probability density, it is straightforward to obtain the expectation and the variance of the process. We get, respectively
\begin{equation}
\label{Xmeantype_alpha_beta*}
	\E[X^*_{\alpha \beta}(t)] =  \beta + \alpha  t - e^{-\frac{\beta ^2}{2 t}}\sqrt{\frac{2 t}{\pi }}  -\frac{\left(\beta ^2+t\right) \erf\left(\frac{\beta }{\sqrt{2 t}}\right)}{\beta } ,
\end{equation}
and
\begin{equation}
\label{Xvartype_alpha_beta*}
	 \Var[X^*_{\alpha \beta}(t)] = \beta ^2+ \left(3-\frac{2 e^{-\frac{\beta ^2}{t}}}{\pi }\right) t - \frac{\left(\beta ^2+t\right) \erf\left(\frac{\beta }{\sqrt{2 t}}\right) \left(\left(\beta ^2+t\right) \erf\left(\frac{\beta }{\sqrt{2 t}}\right) + 2 \beta \sqrt{\frac{2 t}{\pi }} e^{-\frac{\beta ^2}{2 t}}\right)}{\beta ^2} \, ,
\end{equation}
and their asymptotic behaviors for large times
\begin{equation}
\label{Xmean_and_Var_asymptotic_type_alpha_beta*}
   \left\{
       \begin{aligned}
        \E[X^*_{\alpha \beta}(t)]  - \alpha  t    & \underset{t \to \infty}{\sim\,} -2 \sqrt{\frac{2 t}{\pi }}  \\
        \Var[X^*_{\alpha \beta}(t)]     & \underset{t \to \infty}{\sim\,} \left(3-\frac{8}{\pi }\right) t \, ,
       \end{aligned}
   \right.
\end{equation}
which are the same as the pure taboo process Eq.\eqref{Xmean_and_Var_asymptotic_typeI}. Figure~\ref{fig4} shows some realizations of the process $X^*_{\alpha \beta}(t)$.

\subsection{Conditioned process with generator $\mathcal{L}_{e}. = \frac{1}{T-t}\left( X \coth \left(\frac{x X}{T-t}\right) -x \right)  \frac{\partial .}{\partial x}  + \frac{1}{2}  \frac{\partial^2 .}{\partial x^2}$ (generalized Brownian excursion)} 
\label{subsec6}
In this paragraph, we consider a generalized Brownian bridge starting at $X_{e}(0) = x_0$ and ending at $X_{e}(T) = X$ conditioned to stay positive. This process is sometimes considered as a generalized Brownian excursion (recall that a Brownian excursion is a one-dimensional Brownian motion over the time interval $0 \leq t \leq T$ that begins and ends at the origin $x(0) = x(T) =
0$, but is constrained to remain positive in between) and we will use this terminology.\\
It is well-known that conditioning a Brownian motion or a drifted Brownian motion on its final state leads to the same process: the Brownian bridge, which is thus independent of the original drift. Therefore, without loss of generality, in this section we consider a driftless Brownian motion that starts at $B(0) = x_0$ and ends at $B(T) = X$, conditioned to stay positive. To derive the effective drift of this process, we need the conditioned probability defined by
\begin{equation}
\label{conditioned_probability_definition_excursion_dritfless}
	\pi(x,t,X,T) = \mathrm{Pr}[B(u) \geq 0 \mathrm{~~for~all~~}  t \leq u \leq T \vert B(t)=x, B(T)=X] 	\, .
\end{equation}
This conditioned probability $\pi(x,t,X,T)$ is well-known and can be obtained thanks to the method of images~\cite{ref_book_Redner,ref_Monthus_Mazzolo},
\begin{equation}
\label{conditioned_probability_excursion_dritfless}
	\pi(x,t,X,T) =  \frac{1}{\sqrt{2 \pi  (T-t)}}	\left(e^{-\frac{(X-x)^2}{2(T-t)}}  - e^{-\frac{(X+x)^2}{2(T-t)}}   \right)\, .
\end{equation}
The corresponding effective drift of the generalized Brownian excursion is derived from Eq.\eqref{Doob_general} and can be found in~\cite{ref_Majumdar_Orland}
\begin{equation}
\label{mu_excursion_Majumdar_Orland}
	\mu_e(x,t) = \frac{\left( \frac{X-x}{T-t} \right)e^{-\frac{(X-x)^2}{2(T-t)}} +  \left( \frac{X+x}{T-t} \right)e^{-\frac{(X+x)^2}{2(T-t)}}  }{e^{-\frac{(X-x)^2}{2(T-t)}}  - e^{-\frac{(X+x)^2}{2(T-t)}} } \, ,
\end{equation}
or more compactly
\begin{equation}
\label{mu_excursion}
	\mu_e(x,t) = \frac{1}{T-t}\left( X \coth \left(\frac{X x}{T-t}\right) -x \right) \, ,
\end{equation}
which is the drift corresponding to the generator $\mathcal{L}_{e}$. This result can be verified through direct calculation by considering the conditioned probability of the drifted Brownian motion defined by Eq.\eqref{conditioned_probability_definition_excursion} and given by Eq.\eqref{conditioned_probability_excursion} 
\begin{equation}
	\pi(x,t,X,T,\mu) =  \frac{1}{\sqrt{2 \pi  (T-t)}}	\left(e^{-\frac{(X-x-\mu (T-t))^2}{2(T-t)}}  - e^{-2 \mu x} e^{-\frac{(X+x-\mu (T-t))^2}{2(T-t)}}   \right) .
\end{equation}
Applying Doob's transform yields
\begin{equation}
    \mu_{e}(x,t,\mu) = \mu + \frac{1}{\pi(x,t,X,T,\mu)}   \frac{\partial \pi(x,t,X,T,\mu)}{\partial x} = \frac{1}{T-t}\left( X \coth \left(\frac{X x}{T-t}\right) -x \right) \, ,
\end{equation}
an expression that is independent of $\mu$ and corresponds to the effective drift resulting from conditioning a pure Brownian motion given by Eq.\eqref{mu_excursion}. Observe that when $X \to 0$, 
\begin{equation}
\label{mu_excursion_standard}
 \lim_{X \to 0} \mu_e(x,t) = \frac{1}{x} -\frac{x}{T-t}
\end{equation} 
is the drift of the standard Brownian excursion~\cite{ref_book_Rogers_Williams} as it should. From Eq.\eqref{mu_excursion}, it is not obvious to see that the process has an entrance boundary. However, looking at the behavior of the drift at the origin $\mu_e(x,t) \opsim_{x  \to 0}  \frac{1}{x}$ reveals that the origin is indeed an entrance boundary (in fact, for a drift of the form $\alpha/x$, the origin is an entrance boundary when $\alpha \geq 1/2$, see for example references~\cite{ref_Kent} or~\cite{ref_Mazzolo_Taboo}). Now, using the transformation:
\begin{equation}
   W(t) \mapsto \log \left(\sinh \left(\frac{X W(t)}{T-t}\right)\right)-\frac{W(t)^2}{2 (T-t)}  \, ,
\end{equation}
the probability density can be obtained using the technique explained in the previous paragraphs. After some calculations, we obtain the probability density $p_{e}(x,t,x_0)$ of the generalized Brownian excursion
\begin{equation}
\label{drifted_excursion_density}
 	p_{e}(x,t,x_0,X) = \sqrt{\frac{T}{2 \pi t (T-t)}}   \frac{\sinh \left(\frac{x X}{T-t}\right) }{ \sinh\left(\frac{x_0 X }{T}\right)}  e^{-\frac{t X^2}{2 T(T-t)}} \left(e^{-\frac{(t x_0+T (x-x_0))^2}{2 t T (T-t)}}-e^{-\frac{(t x_0-T (x+x_0))^2}{2 t T (T-t)}}\right) \, .
\end{equation}
When the generalized excursion starts at the origin, the expression simplifies to
\begin{equation}
\label{drifted_excursion_density_starts_at_zero}
 	p_{e}(x,t,0,X) =\sqrt{\frac{2 T^3}{\pi t^3(T-t)}}  e^{-\frac{t^2 X^2+T^2 x^2}{2 T t (T-t)}} \frac{x}{ X } \sinh \left(\frac{x X}{T-t}\right) \, .
\end{equation}
Moreover if, $X=0$, then 
\begin{equation}
\label{drifted_excursion_density_starts_and_ends_at_zero}
 	p_{e}(x,t,0,0) = \sqrt{\frac{2 T^3}{\pi t^3 (T-t)^3 }}  x^2 e^{-\frac{T x^2}{2 t (T-t)}} \, ,
\end{equation}
recovering the density of the Brownian excursion~\cite{ref_Takacs}.

\noindent For the sake of completeness, we also report the mean and the variance of generalized Brownian excursion, they read as follows
\begin{equation}
\label{drifted_excursion_mean}
   \E[X_{e}(t)] = \frac{ e^{\frac{X x_0}{T}} (t (X-x_0)+T x_0) \erf\left(\frac{t (X-x_0)+T x_0}{\sqrt{2 t T (T-t)}}\right)-e^{-\frac{X x_0}{T}}  (t (X+x_0)-T x_0) \erf\left(\frac{t (X+x_0)-T x_0}{\sqrt{2 t T (T-t)}}\right)}{2 T \sinh\left(\frac{X x_0}{T}\right)}  \, ,
\end{equation}
and
\begin{align}
\label{drifted_excursion_variance}
   \Var[X_{e}(t)] & = \frac{1}{4 T^2 \sinh^2\left(\frac{X x_0}{T}\right)} \left( 2 (t (X-x_0)+T x_0) (t (X+x_0)-T x_0) \erf\left(\frac{t (X-x_0)+T x_0}{\sqrt{2 t T (T-t)}}\right) \erf\left(\frac{t (X+x_0)-T x_0}{\sqrt{2 t T (T-t)}}\right)  \right. \\
                  &  -e^{\frac{2 X x_0}{T}} (t (X-x_0)+T x_0)^2 \erf\left(\frac{t (X-x_0)+T x_0}{\sqrt{2 t T (T-t)}}\right)^2 - e^{-\frac{2 X x_0}{T}} (T x_0-t (X+x_0))^2 \erf\left(\frac{t (X+x_0)-T x_0}{\sqrt{2 t T (T-t)}}\right)^2   \\
                  & + \left. 4 \left(t \left(t \left(X^2-T\right)+T^2\right)+ (T-t)^2 x_0^2 \right) \sinh ^2\left(\frac{X x_0}{T}\right) - 4 t(T-t) X x_0  \sinh \left(\frac{2 X x_0}{T}\right) \right) \, .
\end{align}
Note that $\Var[X_{e}(0)]=\Var[X_{e}(T)]=0$, as it should, since the generalized Brownian excursion starts at $x_0$ and is pinned down to the value $X$ at time $T$.


\subsection{Conditioned process with generator $\mathcal{L}_{m}. =  \left( \mu + \frac{   2 \sqrt{\frac{2}{\pi (T-t)}} e^{-\frac{(x+\mu  (T-t))^2}{2 (t-T)}}  +2 \mu  e^{-2 \mu  x} \erfc\left(\frac{x - \mu  (T-t)}{\sqrt{2 (T-t)}}\right)}{1 + \erf\left(\frac{x + \mu  (T-t)}{\sqrt{2 (T-t)}}\right)-e^{-2 \mu  x} \erfc\left(\frac{x -\mu (T-t)}{\sqrt{2 (T-t)}}\right)} \right)  \frac{\partial .}{\partial x}  + \frac{1}{2}  \frac{\partial^2 .}{\partial x^2}$ (Drifted Brownian meander)} 
\label{subsec7}
The drifted Brownian meander is a drifted Brownian motion conditioned to remain positive. In the following, the drifted Brownian meander $X_m(t)$ starts at $X_m(0) =  x_0 \geq 0$ and ends at $X_m(T) =  X \geq 0$, while remaining positive in $t \in [0,T]$. To derive the effective drift of the process, we need the conditioned probability defined by
\begin{equation}
\label{conditioned_probability_definition_excursion}
	\pi(x,t,X,T,\mu) = \mathrm{Pr}[B^{\mu}(u) \geq 0 \mathrm{~~for~all~~}  t \leq u \leq T \vert B^{\mu}(t)=x, B^{\mu}(T)=X] 	\, ,
\end{equation}
and since the final position $X$ can be anywhere in the upper half-plane, the above probability must be further integrated for $X \in [0,\infty]$, that is
\begin{equation}
\label{conditioned_probability_definition_meander}
	\pi_m(x,t,T,\mu) = \int_0^{\infty} \pi(x,t,X,T,\mu) dX \, .
\end{equation}
The conditioned probability $\pi(x,t,X,T,\mu)$ is well known and can be obtained thanks to the method of images~\cite{ref_book_Redner,ref_Monthus_Mazzolo}
\begin{equation}
\label{conditioned_probability_excursion}
	\pi(x,t,X,T,\mu) =  \frac{1}{\sqrt{2 \pi  (T-t)}}	\left(e^{-\frac{(X-x-\mu (T-t))^2}{2(T-t)}}  - e^{-2 \mu x} e^{-\frac{(X+x-\mu (T-t))^2}{2(T-t)}}   \right)\, ,
\end{equation}
therefore
\begin{equation}
\label{conditioned_probability_meander}
	\pi_m(x,t,T,\mu) = \frac{1}{2} \left(1  + \erf\left(\frac{x + \mu  (T-t)}{\sqrt{2 (T-t)}}\right)-e^{-2 \mu  x} \erfc\left(\frac{x -\mu (T-t)}{\sqrt{2 (T-t)}}\right)\right) \, ,
\end{equation}
where $\erf(x)= \frac{2}{\pi}\int_0^x e^{-u^2} du $ is the Error function and $\erfc(x) = 1 - \erf(x)$ the complementary Error function. 
The corresponding effective drift of the drifted Brownian meander is derived from Eq.\eqref{Doob_general} and reads
\begin{equation}
\label{mu_drifted_meander}
	\mu_m(x,t) = \mu + \frac{   2 \sqrt{\frac{2}{\pi (T-t)}} e^{-\frac{(x+\mu  (T-t))^2}{2 (t-T)}}  +2 \mu  e^{-2 \mu  x} \erfc\left(\frac{x - \mu  (T-t)}{\sqrt{2 (T-t)}}\right)}{1 + \erf\left(\frac{x + \mu  (T-t)}{\sqrt{2 (T-t)}}\right)-e^{-2 \mu  x} \erfc\left(\frac{x -\mu (T-t)}{\sqrt{2 (T-t)}}\right)} \, ,
\end{equation}
which is the drift corresponding to the generator $\mathcal{L}_{m}$. It is not obvious from Eq.\eqref{mu_drifted_meander} that the process has an entrance boundary (at the origin). However, by looking at the behavior of the drift at the origin
\begin{equation}
\label{mu_drifted_meander_at_the_origin}
	\mu_m(x,t) \opsim_{x  \to 0} \mu + \frac{1}{x} \, ,
\end{equation}
we can see that the origin is indeed an entrance boundary. Now, with the help of the transformation: 
\begin{equation}
W(t) \mapsto \mu  W(t) + \log \left[\frac{1}{2} \left(1 + \erf\left(\frac{W(t) + \mu  (T-t)}{\sqrt{2(T-t)}}\right)-e^{-2 \mu  W(t)} \erfc\left(\frac{W(t) -\mu (T-t)}{\sqrt{2(T-t)}}\right)\right)\right] \, ,
\end{equation}
the probability density can be obtained using the technique described in the previous paragraphs. After some calculations, we obtain the explicit expression of the probability density $p_{m}^{\mu}(x,t,x_0)$ of the generalized Brownian meander
\begin{equation}
\label{drifted_meander_density}
	 p_{m}^{\mu}(x,t,x_0) = \frac{\left(e^{\frac{2 x x_0}{t}}-1\right) e^{-\frac{2 x (x_0+\mu  t)+(x_0-\mu  t)^2+x^2}{2 t}} \left(\erf\left(\frac{x -\mu (T-t)}{\sqrt{2 (T-t)}}\right)+e^{2 \mu  x} \left(\erf\left(\frac{x + \mu (T-t)}{\sqrt{2 (T-t)}}\right)+1\right)-1\right)}{\sqrt{2 \pi t} \left(e^{2 \mu x_0} \left(\erf\left(\frac{x_0 + \mu  T}{\sqrt{2 T}}\right)+1\right)+\erf\left(\frac{x_0-\mu  T}{\sqrt{2 T}}\right)-1\right)} .
\end{equation}
When $x_0 = 0$ the preceding equation simplifies to
\begin{equation}
\label{drifted_meander_density_zero}
	 p_{m}^{\mu}(x,t,0) =  \frac{\sqrt{T} x e^{\frac{1}{2} \left(\mu ^2 (T-t)-\frac{x^2}{t}-2 \mu  x\right)} \left(\erf \left(\frac{x - \mu  (T-t)}{\sqrt{2 (T-t)}}\right)+e^{2 \mu  x} \left(\erf\left(\frac{x +\mu  (T-t)}{\sqrt{2 (T-t)}}\right)+1\right)-1\right)}{t^{3/2} \left(\mu \sqrt{2 \pi T} e^{\frac{\mu ^2 T}{2}} \left(\erf\left( \mu \sqrt{\frac{T}{2}} \right)+1\right)+2\right)}.
\end{equation}
When $\mu = 0$, from Eq.\eqref{drifted_meander_density} the density of the Brownian meander reads
\begin{equation}
\label{meander_density}
 p_{m}^{0}(x,t,x_0) = \frac{e^{-\frac{(x+x_0)^2}{2 t}} \left(e^{\frac{2 x x_0}{t}}-1\right) \erf\left(\frac{x}{\sqrt{2 (T-t)}}\right)}{\sqrt{2 \pi t} \erf\left(\frac{x_0}{\sqrt{2 T}}\right)}.
\end{equation}
Besides, when $x_0 =0$, the density reduces to
\begin{equation}
\label{meander_density_zero}
	 p_{m}^{0}(x,t,0) = \frac{\sqrt{T} x e^{-\frac{x^2}{2 t}} \erf\left(\frac{x}{\sqrt{2 (T-t)}}\right)}{t^{3/2}} \, ,
\end{equation}
and at the final time $T$
\begin{equation}
\label{meander_density_zero_at_T}
	 p_{m}^{0}(x,T,0) = \frac{x e^{-\frac{x^2}{2 T}}}{T} \, ,
\end{equation}
recovering the famous truncated Rayleigh distribution~\cite{ref_Iafrate}.

\section{Conclusion}
\label{sec_Conclusion}
In this article we have derived the closed-form of the probability density of several conditioned processes with an entrance boundary. Due to this entrance boundary, the drifts have singularities and standard methods for obtaining exact results are usually doomed to fail. To overcome this difficulty, we have developed an original and robust method that combines Girsanov's theorem and a variant of the image method. We have applied this general framework to the cases of Brownian motion conditioned to remain below a line $\beta + \alpha t$, where $\beta >0$. Depending on the sign of $\alpha$, the conditioned processes can behave very differently. If $\alpha$ is positive, the state space expands, and eventually the process no longer feels the repulsive barrier and evolves almost freely. On the other hand, if $\alpha$ is negative, the state space shrinks and the process is affected at all times by the repulsive barrier, which strongly influences its average behavior. Moreover, the method allows us to derive the exact probability density of the generalized Brownian excursion and the drifted Brownian meander, recently introduced in~\cite{ref_Iafrate}.






\end{document}